\documentstyle[twoside,fleqn,espcrc2]{article}

\newcommand{\AmS}{{\protect\the\textfont2
  A\kern-.1667em\lower.5ex\hbox{M}\kern-.125emS}}

\title{
Instanton-induced production of jets with large transverse momentum
 in QCD  }
\author{I.I. Balitsky\address{
        Physics Department, Penn State University,
       104 Davey Lab.,University Park, PA 16802, USA}%
        \thanks{ On leave of absence from
St.Petersburg Nuclear Physics Institute, 188350 Gatchina, Russia}
        and
  V.M. Braun\address{Max-Planck-Institut f\"ur Physik,
                   P.O.Box 40 12 12, Munich, Germany}$^\ast$}

\begin{document}

\begin{abstract}
We consider the instanton-induced cross section for production of a
gluon jet with large transverse momentum in QCD and
point out that Mueller's corrections corresponding to the
rescattering of hard quanta are likely to remove contributions of
large instantons, making this cross section well defined.
Some speculations about possible phenomenological signatures are
presented.
\end{abstract}

\maketitle


In recent years there is a revival of interest to instanton
effects in gauge theories, inspired by the conjection \cite{ring90}
that in high-energy collisions such effects are enhanced enough to
produce an observable violation of the baryon number.
Although present theoretical arguments rather seem to disfavour such
a strong enhancement \cite{matt92}, so that
the instanton effects in the electroweak theory presumably
remain far below the level at which
they might become observable at future colliders, these studies
have trigged an increasing  interest to semiclassical
effects in gauge theories at high energies in a more general context.

In \cite{bal93a,bal93b} we have suggested to look for the
instanton contributions in high-energy collisions in the QCD.
In this case the coupling is not so small as in the electroweak
theory, and the instanton contributions might be observable
even if they remain exponentially supressed. The Ringwald's
phenomenon --- enhancement of the cross section through the
dominance of final states with many gluons ---
allows one to hope that these cross sections may reach observable
values, and similtaneously provides a good trigger for their
observation, since a fireball of $\sim 2\pi/\alpha_s$ gluons
is likely to produce an event with a very high density of particles
in the final state.

The major difficulty for the identification of instanton
effects in QCD is that in the generic situation they are not
infra-red (IR) stable, typically involving
a IR-divergent integral over the instanton sizes.
There are speculations that in the true QCD vacuum this
integration is effectively cut off at sizes of order $\rho\sim
1/600$ MeV, but this assertion is difficult to justify
theoretically.

To be on  a safer side, one should take special care
to select contributions of small instantons, which implies
going over to a certain hard process.
This notion is applied to
reactions in which hard scale is related either
to a large virtuality
of the external particle (photon or W,Z boson),  or
to a large momentum transfer.
 In perturbation theory this
distinction is subtle: a large momentum transfer necessarily
involves an exchange by a highly-virtual gluon (quark).
Thus, perturbative description of these processes
is similar, in both cases
dynamics of small distances can be factorized from
the large distance effects. It is this way that
most of the QCD predictions arise.

This distinction is crucial, however, for the discussion of
instanton effects.
In the deep inelastic lepton-hadron
scattering the
hard scale is brought in by the virtuality $Q^2$ of the photon.
In this case \cite{bal93a,bal93b} the
contribution of small instantons is distinguished
by a non-trivial power dependence on $Q$, corresponding
to a fractional twist, and can be
disentangled from IR divergent contributions
of large instantons, included in parton distributions.  Thus,
coefficient functions in front of parton distributions receive
well-defined non-perturbative corrections
\begin{equation}\label{CF}
C(x,Q^2)\sim
e^{-\frac{4\pi}{\alpha_s(\rho)}
\left[1-\frac{3}{2}\left(\frac{1-x}{1+x}\right)^2+\ldots\right]
}\,,
\end{equation}
coming from instantons with the size of order
 $           \rho \sim \pi/\alpha_s\cdot{1}/{Q}$,
which turn out to be of order $10^{-2}-10^{-5}$
in the region of sufficiently large $Q^2>100$ GeV$^2$
and Bjorken variable $x>0.3$, where the derivation of (\ref{CF})
is justified.

The situation proves to be essentially different in large
momentum transfer reactions, from which we consider
production of a gluon jet with large $q_\perp$    as a
representative example. On physical grounds it
is obvious that this cross section cannot be affected by large
instantons.
However,
a semiclassical calculation
fails in this case: to the accuracy to which (\ref{CF}) is derived,
the instanton contribution is given by a power-like divergent
integral, and contributions of small instantons $\rho\sim 1/q_\perp$
do not produce any non-trivial dependence on $q_\perp$.
Indeed, to the semiclassical accuracy the effect of small instantons
is to introduce
new {\em point-like } multi-particle vertices, which do not
involve
any momentum transfer dependence. Thus, instanton-induced amplitudes
do not decrease at large momentum transfers.

Conceptually, it is easy to realize what is missing: to obtain a
sensible result one
must take into account an (exponentially small) overlap
between the initial state, which involves a few hard quanta, with the
semiclassical final state \cite{banks}.
This necessarily involves taking into account quantum corrections
to semiclassical amplitudes in the instanton background, the study
of which has been pioneered by Mueller \cite{MU91}, see \cite{matt92}
for a review and further references.
We demonstrate that the ``Mueller's corrections" indeed
remove contributions of large instantons to the
jet production with large $q_\perp$, making the
non-perturbative contribution to this cross section well defined.


Mueller finds \cite{MU91} that to the one-loop accuracy the
asymptotics of the gluon propagator in the instanton background
takes the factorized form
\begin{equation}\label{propagator}
   G(p,q) \simeq A_I(p) A_I(q) \frac{\alpha_s}{8\pi}(pq)\ln(pq)
\,,
\end{equation}
assuming $pq\gg 1/\rho^2$, $p^2=q^2=0$.
{}From this, it is possible to derive that the corresponding quantum
correction to the instanton-induced amplitude acquires the factor
\begin{equation}\label{Mgeneral}
   \exp\left\{-\frac{\alpha_s}{8\pi}M(p_i)\right\}\,,
\end{equation}
where
\begin{equation}\label{Msum}
 M(p_i) = \sum_{i<j} (p_i\cdot p_j) \ln (p_i\cdot p_j)\,.
\end{equation}
The summation goes over all the ingoing and outgoing particles.
To $O(\alpha_s)$ accuracy this formula follows directly
from (\ref{propagator}), provided $(p_i\cdot p_j)\gg 1/\rho^2$,
while the
exponentiation of this result is a plausible
conjecture beyond, see \cite{matt92}.
Our strategy here is to take the quantum correction in (\ref{Mgeneral})
for granted, and evaluate its effect on the jet production.
We note in passing that in the deep inelastic scattering
the quantum correction (\ref{Mgeneral}) is
necessary to cancel the ambiguity in the $\bar I I$ interaction.
In this case the instanton size determined from the saddle-point
equations is of order
$\rho\sim(4\pi)/\alpha_s\cdot 1/Q\cdot 1/\xi^2$ where
$\xi\simeq(R/\rho)^2$ \cite{bal93a}, and the $\bar I I$ separation
$R$ is fixed by the Bjorken $x$ alias by the initial energy.
At large $\xi\gg 1$ one has, generally, $M\sim Q^2$ and thus the
expression under the exponent in (\ref{Mgeneral}) is of order
$\pi/\alpha_s \cdot 1/\xi^4$, same as ambiguities in
$U^{\bar I I}_{\rm int}$.

 The Mueller's factor $M$ (\ref{Msum}) involves momenta
of hard particles (a few), which we denote by $p_i$, and soft
particles ($n\sim (4\pi)/\alpha_sU^{\bar I I}_{\rm int}$),
denoted by $k_i$.
Hard particles are the two colliding partons (gluons) and the
final state gluons with large momentum, which are resolved as
jets. We consider the inclusive cross section, summing over
all soft particles and integrating over their
phase space.

It is possible to prove that at large $\xi$
the soft momenta appearing in $M$ can be
substituted by
\begin{equation}\label{sub}
     k_i \rightarrow \frac{E=\sum k_i}{n}\,.
\end{equation}
The derivation
of (\ref{sub}) will be given in \cite{bal94}.
The dependence on $n$ actually cancels, and $M$
is expressed to this accuracy entirely in
terms of hard momenta. For the back-to-back production of a
pair of gluon jets $ g+g \rightarrow g+g+X$  we find
\begin{equation}
   M= 2 \ln 2 (p-q)^2 + 4 pq \ln 2 \,G(\theta)\,,
\end{equation}
\begin{equation}
\ln 2\,
G(\theta) = -\sin^2\frac{\theta}{2} \ln\sin^2\frac{\theta}{2}
    - \cos^2\frac{\theta}{2} \ln\cos^2\frac{\theta}{2}\,,
\end{equation}
where $p,q$ are the momenta of ingoing and outgoing gluons in c.m.
frame,
respectively, and $\theta$ is the angle between them.
The instanton-induced cross section to the exponential accuracy
reads
\begin{equation} \label{sigma}
\sigma_I\sim \int dR\,d\rho\, e^{ER-\frac{4\pi}{\alpha_s} S(\xi)
 -\frac{\alpha_s}{4\pi}M(p,q)\rho^2}\,.
\end{equation}
Here $S(\xi)$ is the QCD action on the $\bar I I$ configuration,
and $E=2(p-q)$ is the energy transferred to the instanton
(and released in soft particle emission).
The integral is taken by the saddle-point method. Neglecting for
simplicity the running of the QCD coupling and taking into account
the dipole interaction term in the expansion of the action
$S(\xi) = 1-6/\xi^2+\ldots$, one gets the saddle-point values for
$\rho$ and $R$ from the equations
\begin{equation}\label{saddle}
 \frac{\alpha_s}{\pi} \rho = \frac{2E}{M}\sqrt{\xi}\,,~~
 \frac{E^2}{M} = \frac{48}{\xi^3}\,.
\end{equation}
Now comes the central point. The function $G(\theta)$ varies
between 0 and 1, with a minimum value $G=0$ at $\theta=0$ and
$\theta=\pi$, and a maximum $G=1$ at $\theta=\pi/2$.
Consider first the collinear jet production, $\theta=0,\pi$.
Hence $M=2 \ln 2 (p-q)^2 = E^2/2\cdot\ln 2$ is of order of the energy
transferred to the instanton. From the second of the saddle-point
equations in (\ref{saddle}) one finds then $\xi^3 = 24 \ln 2$,
independent on the external momenta. Thus, in this
case the cross section is defined by the region of $R\sim \rho$ where
instantons interact strongly and the calculation is not justified
(parametrically). On the other hand, consider jets with
large transverse momentum, $\theta = \pi/2$. Then $M=2\ln 2 (p^2+q^2)
\simeq s \ln 2$, where $s=4 p^2$ is the total energy, and
substituting this to the saddle-point equation we find
$   \xi^3 = 48 \ln 2 \cdot s/E^2$.
Keeping $E^2\ll s$ (which means that momenta of gluon jets are
close to momenta of coliding gluons) we get $R\gg\rho$ and
the calculation is under control.
Note that we get $\alpha_s/4\pi\cdot \rho^2 M \sim \pi/\alpha_s
\cdot 1/\xi^2$, which indicates that the Mueller's correction
now contributes on the leading $1/\xi^2$ level, same as the
dipole interaction.

The same effect is observed in the instanton-induced production
of monojets $g+g\rightarrow g+X$, in which case for $\theta=\pi/2$
in the c.m. frame we obtain
\begin{eqnarray}
M &=& s \ln 2 -(s/2) \ln(1+E^2/s)+
(E^2/2) \ln 2 \nonumber\\ &&{}
- (E^2/2)\ln(1+s/E^2)\,.
\end{eqnarray}
The cross section is again given by the integral in (\ref{sigma}),
and the saddle-point values in the limit $E^2/s \ll 1$ are
\begin{equation}
   \xi^3 = 48 \ln 2\, s/E^2 \,,~~
   \sqrt{s}\,\rho  =
    \frac{4\pi}{\alpha_s}\cdot \frac{\sqrt{3\ln 2}}{\xi}\,.
\end{equation}
 Thus, at least in this academic limit, the calculation is
under control. This example can be interesting phenomenologically,
since it has a clear signature and smaller perturbative background.
Typical numbers are as follows: in gluon-gluon collisions with
$\sqrt{s}=400$ GeV, one could look for production of a monojet
with $q_\perp>180$ GeV, balanced by $n_g\sim 10-15$ gluons and
$2n_f$ quarks with transverse momenta of order $\rho^{-1}\sim
10$ GeV each. The cross section is difficult to estimate, but
is expected to be of the same order or larger than in the deep
inelastic scattering \cite{bal93b}.

To summarise, we have shown that Mueller's corrections are likely
to cut off the IR divergent integrals over the instanton size
in the process of gluon jet production with large transverse
momentum, indicating that their role is more important than
 usually believed.
In  general, one may thus conjecture that in {\em  any} hard
process there is a well-defined nonperturbative contribution due
to small instantons with the size of order $\rho\sim 1/(Q\alpha_s(Q)$,
where $Q$ is the corresponding hard scale.
A search for instanton-induced effects in large $q_\perp$ reactions
may be most fruitful because of larger rates and smaller
backgrounds.

\end{document}